\documentclass[twocolumn]{aastex62}

\graphicspath{{./}{figures/}}
\received{XXX}
\revised{XXX}
\accepted{XXX}
\submitjournal{ApJL}
\shorttitle{Structure in the disc of CI Tau}
\shortauthors{Clarke et al.}
\usepackage{natbib}
\usepackage{bm}

\usepackage{url}
\usepackage{xcolor}
\usepackage{txfonts}
\usepackage{soul}
\usepackage{xspace}
\newcommand{\Mjupiter}{\,M_\mathrm{Jupiter}}

\newcommand{\Jybeam}{Jy\,beam$^{-1}$\xspace}
\newcommand{\muJybeam}{$\mu$Jy\,beam$^{-1}$\xspace}

\usepackage{comment}

\begin{document}
\title{HIGH RESOLUTION MILLIMETRE IMAGING OF THE CI TAU PROTOPLANETARY DISC - A MASSIVE ENSEMBLE OF PROTOPLANETS FROM 0.1 - 100 AU}
\correspondingauthor{Cathie Clarke}
\email{cclarke@ast.cam.ac.uk}
\author{C. J. Clarke}
\affiliation{Institute of Astronomy, Madingley Road, Cambridge, CB3 OHA, UK}
\author{M. Tazzari}
\affiliation{Institute of Astronomy, Madingley Road, Cambridge, CB3 OHA, UK}
\author{A. Juhasz}
\affiliation{Institute of Astronomy, Madingley Road, Cambridge, CB3 OHA, UK}
\author{G. Rosotti}
\affiliation{Institute of Astronomy, Madingley Road, Cambridge, CB3 OHA, UK}
\author{R. Booth}
\affiliation{Institute of Astronomy, Madingley Road, Cambridge, CB3 OHA, UK}
\author{S. Facchini}
\affiliation{Max-Planck-Institut f\"ur Extraterristische Physik, Giessenbachstrasse 1, 85748 Garching, DE}
\author{J.D. Ilee}
\affiliation{Institute of Astronomy, Madingley Road, Cambridge, CB3 OHA, UK}
\author{C.M. Johns-Krull}
\affiliation{Physics \& Astronomy Dept., Rice University, 6100 Main St., Houston, TX 77005, USA}
\author{M. Kama}
\affiliation{Institute of Astronomy, Madingley Road, Cambridge, CB3 OHA, UK}
\author{F. Meru}
\affiliation{Institute of Astronomy, Madingley Road, Cambridge, CB3 OHA, UK}
\affiliation{Department of Physics, University of Warwick, Gibbet Hill Road, Coventry,CV4 7AL, UK}
\affiliation{Centre for Exoplanets and Habitability, University of Warwick, Gibbet Hill Road, Coventry, CV4 7AL, UK}
\author{L.Prato}
\affiliation{Lowell Observatory, 1400 West Mars Hill Road, Flagstaff, AZ 86001 USA}
\begin{abstract}
 
 We present high resolution millimeter continuum imaging of the disc surrounding the young star CI~Tau, a system hosting  the first hot Jupiter candidate in a protoplanetary disc system. The system has extended mm emission on which are superposed three prominent annular gaps at radii $\sim$ 13, 39 and 100\,au. We argue that these gaps are most likely to be generated by massive planets so that, including the  hot Jupiter, the system contains four  gas giant planets at an age of only 2\,Myr. Two of the new planets are similarly located to those inferred in the famous HL Tau protoplanetary disc; in CI Tau, additional observational data enables a  more complete analysis of the system properties than was possible for HL Tau. 
 Our dust and gas dynamical modeling satisfies   every available observational constraint and points to  the most massive  ensemble of exo-planets ever detected at this age, with its four planets spanning a factor  1000 in orbital radius.Our results show that the  association between hot Jupiters and gas giants on wider orbits, observed in older stars, is apparently  in place at an early evolutionary stage.
\end{abstract}
\section{Introduction} \label{sec:intro}
Since the 1995 discovery  of the first hot Jupiter 
 \citep{Mayor}, it is now established that such 
gas giant planets orbiting at radii  $<0.1$\,au from their parent stars are found in around 1\% of main sequence solar type stars \citep{Wright}. There is considerable debate as to whether these objects formed {\it in situ} or have instead migrated from larger radii, either from interaction with their natal protoplanetary disc \citep{Kley} or  planet-planet scattering after the disc has dispersed \citep{Rasio}. With typical ages of up to several Gyr, most hot Jupiter hosts  have long since lost their protoplanetary discs (typical lifetime of a few Myr; \citet{Haisch}); arguments about the origin of hot Jupiters are  thus usually  based on theoretical models  linking hypothetical  initial conditions to present day orbital parameters.
   The recent discovery \citep{Johns-Krull,Biddle}, using the radial velocity technique,  of a hot Jupiter  in the young disc bearing solar type star CI~Tau,  has demonstrated that in at least this case the hot Jupiter is already in a very  close orbit when the star is only $\sim2$\,Myr old \citep{Guilloteau14}.
  CI~Tau is a well studied system, with mass $0.92 M_\odot$ \citep{Simon17}, luminosity $0.93 L_\odot$ \citep{Guilloteau14}, and  is already known to host a massive dust and gas disc extending  many hundreds of\,au from the star \citep{ Guilloteau2011,Andrews07}; additionally it  displays  a high accretion rate of gas  onto the star \citep{McClure13}.
   The  hot Jupiter's mass is $\sim 11.3 \Mjupiter$ if its orbit is aligned with the outer disc 
\citep{Guilloteau14}, consistent with the orbital alignment between hot Jupiters and outer planets found in mature exoplanetary systems \citep{Becker}.
This mass places it in the top 5 \% of the main sequence hot Jupiter population.
   Around half of {\it mature} hot Jupiter systems  also contain companions \citep{Knutson2014,Ngo2015} at less than 20 \,au which, if present at early times, would create structure in the protoplanetary disc. Although
   previous sub-millimetre observations have hinted at  a possible gap in the disc around CI Tau at 100\,au, they  lacked the resolution to characterise this in detail or probe the inner disc where companions may be expected \citep{Konishi2018}.
Here we present high resolution $1.3$mm  ALMA imaging of the disc surrounding CI Tau and  report  three pronounced annular gaps in emission between $10$ and $100$ AU.   
  \begin{figure*}
\centering
\gridline{
\fig{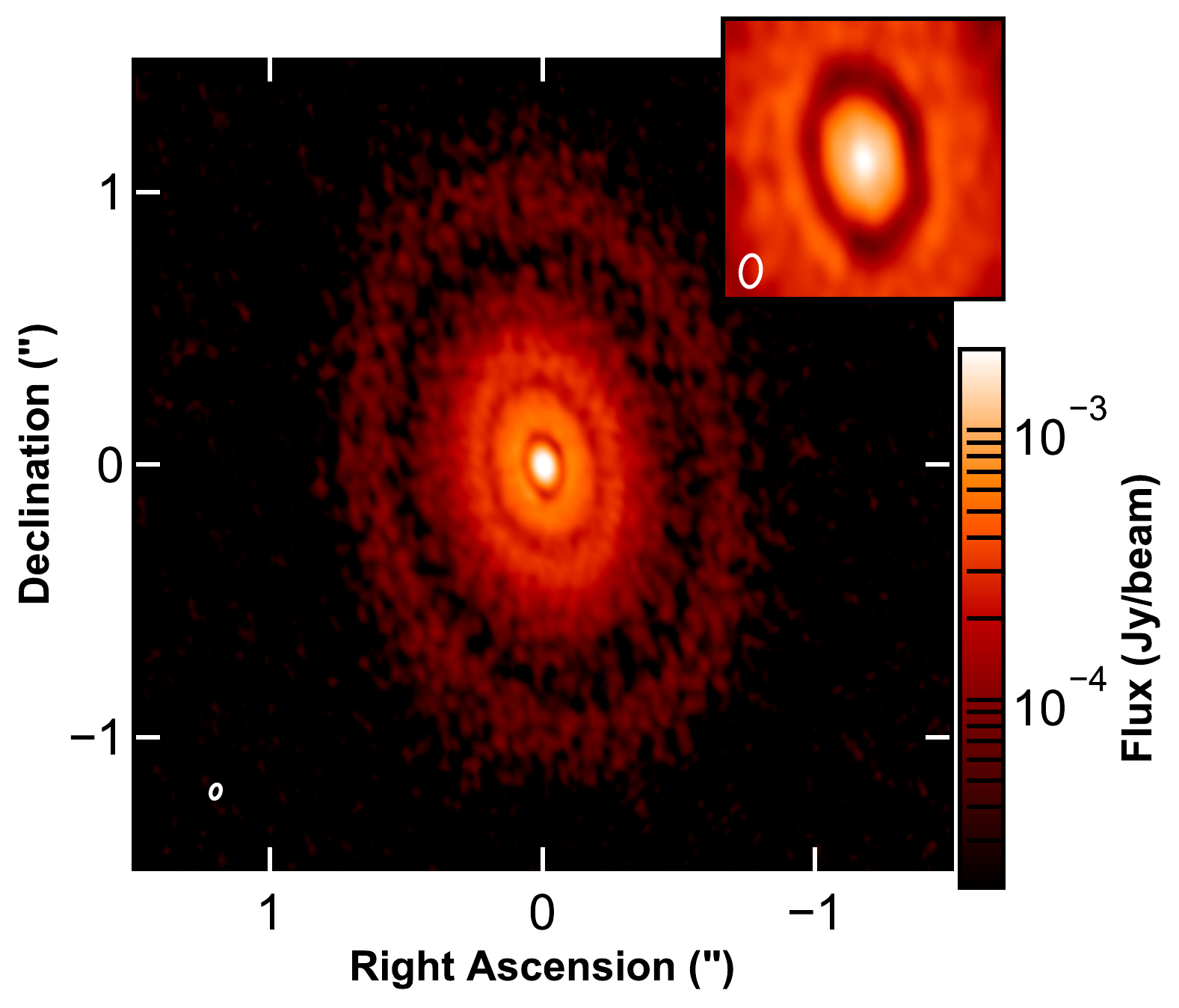}{0.54\textwidth}{\large(a)}
\fig{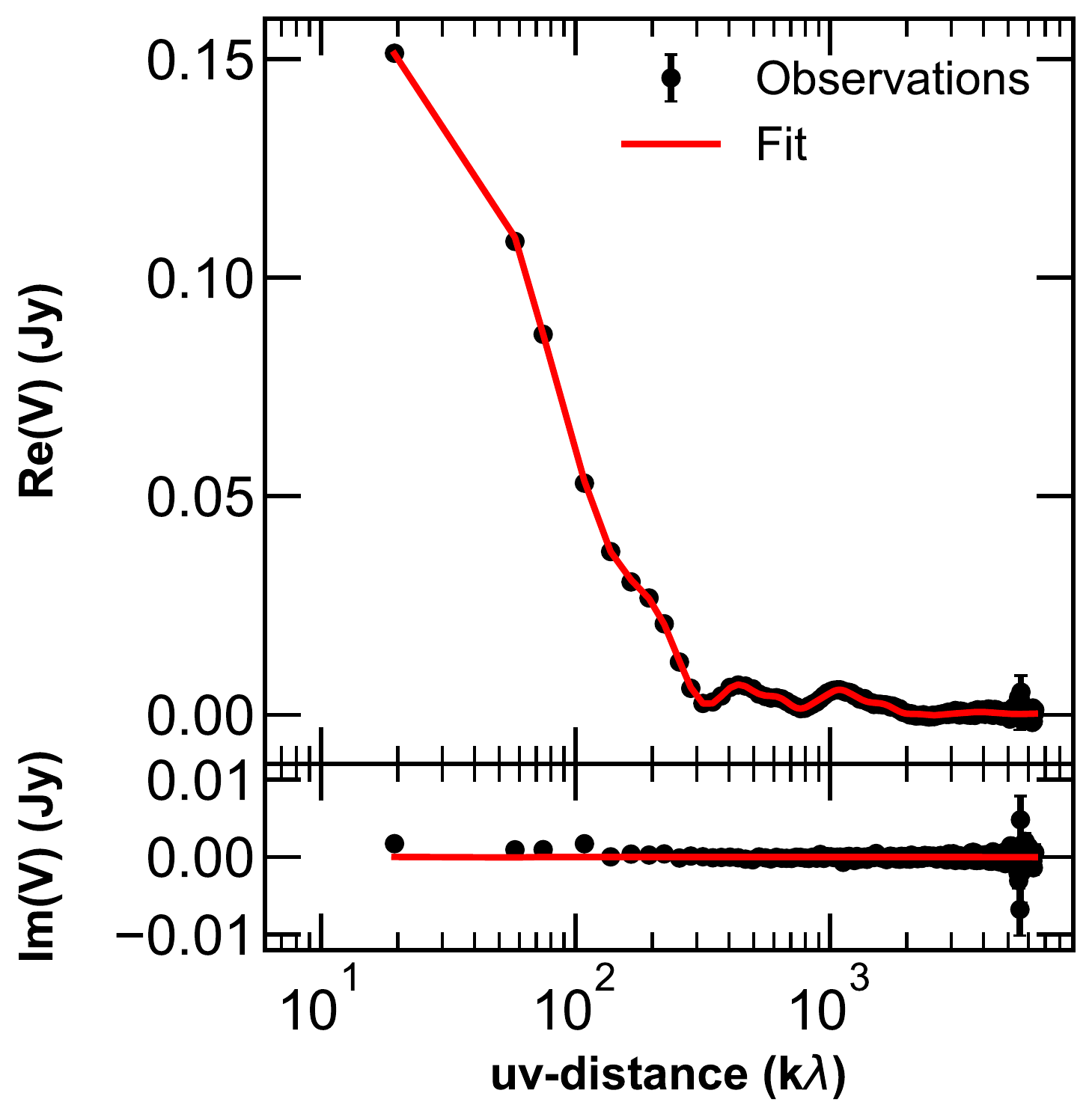}{0.45\textwidth}{\large(b)}}
\gridline{
\fig{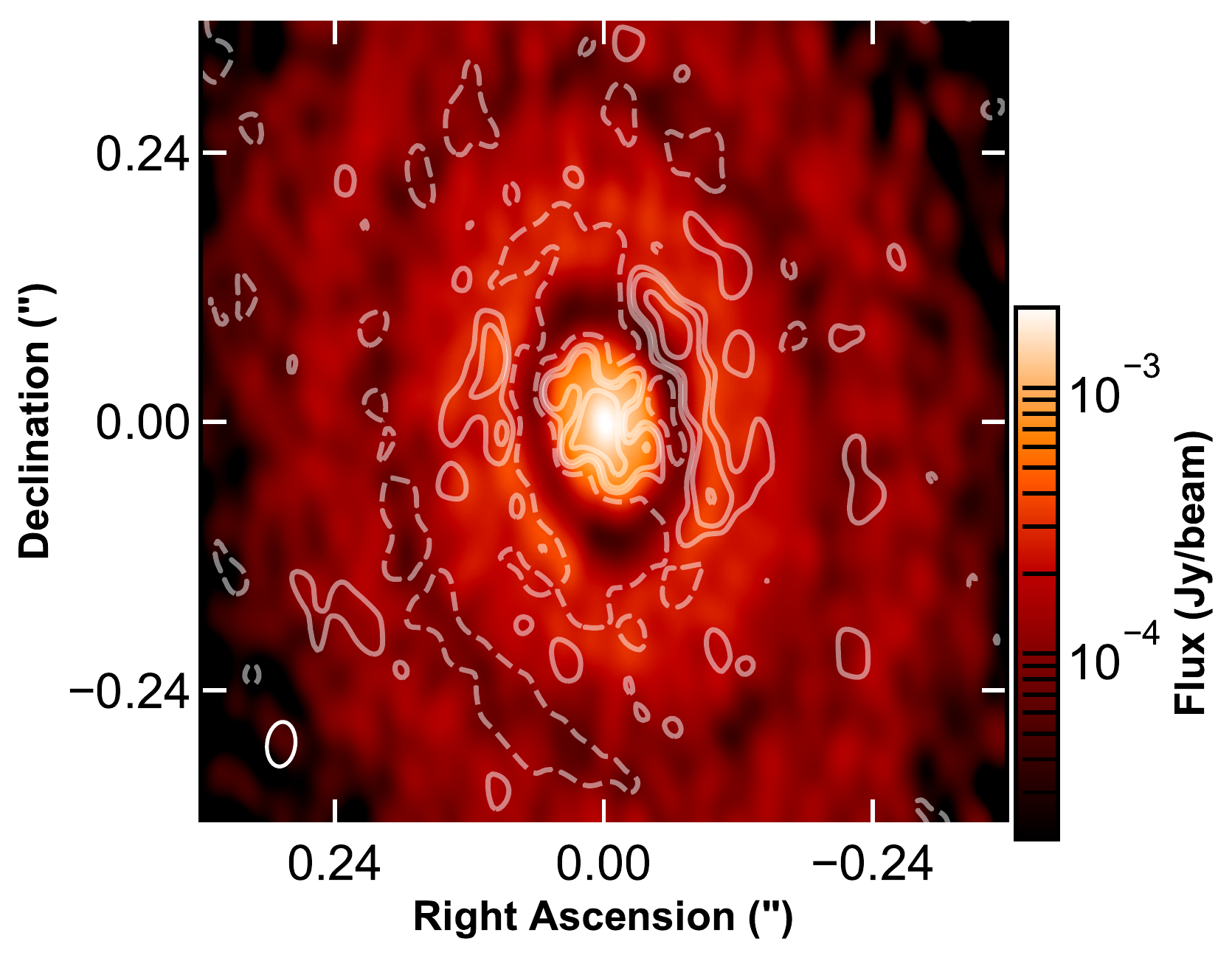}{0.54\textwidth}{\large(c)}
\fig{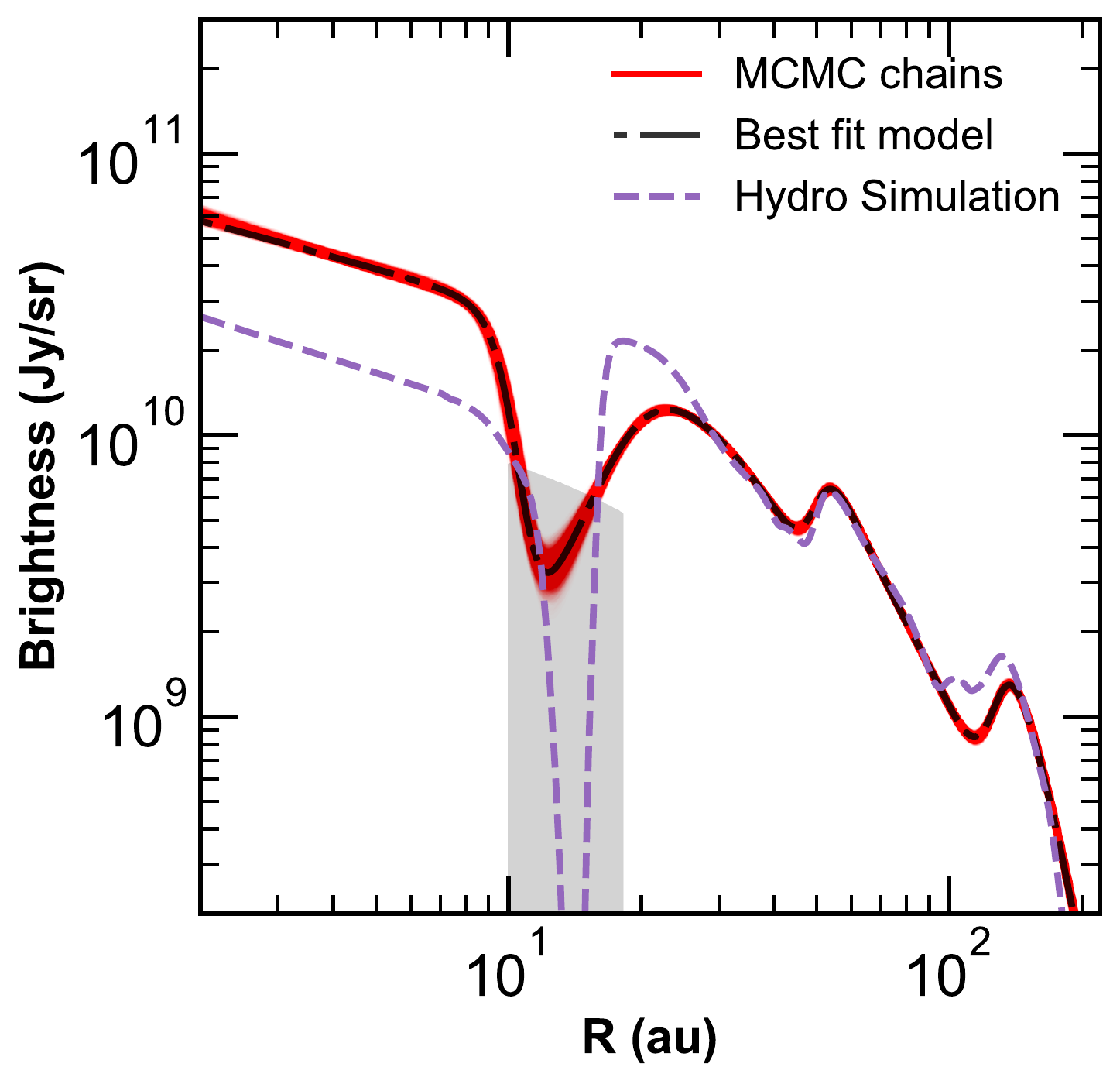}{0.45\textwidth}{\large(d)}
}
\caption{
{(a)} Synthesized image of the CI~Tau continuum observations (beam 50$\times$30 mas FWHM, corresponding to 7$\times$4 au). The rms noise level is $\sigma=13\mu$\Jybeam. The inset shows a 0.35" wide zoom on the innermost gap imaged with a finer resolution (uniform weighting; 40$\times25$ mas or 5$\times$3.5 au, FWHM beam).
{(b)} Observed visibilities compared with the best fit model visibilities as a function of deprojected baseline.
{(c)} Synthesized image (uniform weighting) with fit residuals as white contours drawn at -3$\sigma$, 3$\sigma$, 6$\sigma$, 12$\sigma$, etc.
{(d)} A family of $5\times 10^3$ emissivity profiles drawn from the posterior (red). The black dot-dashed line highlights the best fit model. The gray shaded region indicates the range of gap depths that are still compatible with the observations. The dashed purple line shows the brightness profile obtained from the hydrodynamical simulation.}
  \label{fig:main}
\end{figure*}
We present visibility modeling and  explore the origin of these structures via hydrodynamical modeling, arguing for the presence of three gas giants as outer companions to the hot Jupiter. 
\section{Observations}
CI~Tau was observed with the Atacama Large Millimeter Array (ALMA) on the 23rd and 24th September 2017 (Project ID: 2016.1.01370.S, PI: Clarke) with 40 antennas (baselines between 21 and 12145.2\,m) and an
on-source integration time of 32.35 min in both cases. The correlator  used four spectral windows centered at 224, 226, 240, and 242\,GHz in time division mode to measure the continuum in Band~6.
Each spectral window used 128 channels and a bandwidth of 1.875\,GHz, together providing an effective total continuum bandwidth of 7.5 GHz. To calibrate the visibilities a set of standard calibrators were also observed.  Calibration of the complex interferometric visibilities used the Common Astronomy Software Applications (CASA) v5.1.1 and the ALMA Pipeline.
In panel (a) of Figure~\ref{fig:main} we imaged the calibrated visibilities using the multi-scale CLEAN algorithm with scale parameters of 0", 0.048", 0.08", 0.24" and  Briggs weighting with a robust parameter of 0.5 to obtain the optimal signal-to-noise ratio and spatial resolution. The resulting synthesized beam is $0.05"\times0.03"$ with a position angle of 16 degrees, while the achieved RMS noise level is 13\muJybeam.
\section{Visibility modelling}
We characterise the CI~Tau brightness by fitting the continuum visibilities with an axisymmetric parametric model consisting of an envelope and three gaps. 
For the envelope we use an exponentially-tapered power-law:
\begin{equation}
\label{eq:def.envelope}
I_\mathrm{env}(R)=I_0\left(\frac{R}{R_c}\right)^{\gamma_1}\exp\left[-\left(\frac{R}{R_c}\right)^{\gamma_2}\right]\quad\mathrm{for}\quad{}R\leq R_\mathrm{out}
\end{equation}
with $I_0$  a 
brightness normalisation constant ( Jy sr$^{-1}$). 
Each gap is parametrised as the difference of two logistic functions:
\begin{equation}
\label{eq:def.gap}
I_\mathrm{gap}(R)=\delta_\mathrm{gap}\left[\frac{1}{1+e^{-k_2(R-R_\mathrm{g}-w_2)}}-\frac{1}{1+e^{-k_1(R-R_\mathrm{g}+w_1)}}\right]\,,
\end{equation}
where $\delta_\mathrm{gap}$ 
describes the gap depth ($\delta_\mathrm{gap}=0$ corresponding to no gap), $R_\mathrm{g}$ is the gap radial location, $w_1$ and $w_2$ are the left- and right-hand  gap widths at half depth and $k_1,\ k_2$ express the steepness of the left and right gap profile.
The brightness profile is given by:
\begin{equation}
\label{eq:brightness.profile}
I(R)=I_\mathrm{env}(1+I_\mathrm{gap,1})(1+I_\mathrm{gap,2})(1+I_\mathrm{gap,3})\,,
\end{equation}
involving 5 free parameters for $I_\mathrm{env}$ and 6 free parameters for each $I_\mathrm{gap}$.  We simultaneously fit  the disc inclination $i$ and position angle $PA$ (defined East of North) and the offset $(\Delta$RA,$\Delta$Dec$)$ from the phase center.
We thus have a parameter set $\theta=(I_0,R_c,\gamma_1,\gamma_2,\dots)$ described by 23 parameters for the brightness profile plus 4 for the system geometry.
The 
computation of the visibilities $V_\mathrm{mod}$ for each model $
{\theta}$ 
is performed using GALARIO\footnote{
\href{https://github.com/mtazzari/galario}{https://github.com/mtazzari/galario}.} \citep{Tazzari}, which first computes the 2D image of the disc for a given $I(R)$
and then Fourier transforms and samples it in the observed $(u,v)$ points.
The 
likelihood of the observations $V_\mathrm{obs}$ given the model visibilities $V_\mathrm{mod}(
\theta
)$ is assumed Gaussian:
\begin{equation}
\log{}p(V_\mathrm{obs}\,|\,\theta)=-\frac{1}{2}\,\chi^2=-\frac{1}{2}\sum_{j=1}^{N}|V_\mathrm{mod}-V_\mathrm{obs}|^2w_j\,,
\end{equation}
where $N$ is the total number of visibility points and $w_j$ is the weight\footnote{
The visibility weights $w_j$ are the theoretical estimates obtained by the CASA software package.
} of the $j-$th visibility. The parameter space is explored with a Bayesian approach using the \textsc{emcee} Markov chain Monte Carlo (MCMC) ensemble sampler \citep{Foreman},  providing  an estimate of the posterior probability distribution of the model parameters given the observations:
\begin{equation}
\log{}p(\theta\,|\,V_\mathrm{obs}){}={}\log{}p(V_\mathrm{obs}\,|\,\theta)+\log{}p(\theta){}+{}C\,,
\end{equation}
where $p(\theta)$ is the prior on the parameters and $C$ is a normalisation constant.
Since the parameters are independent, the priors can be written as $p(\theta)=\prod_ip(\theta_i)$.
 We choose uniform priors on all parameters, 
except for  inclination for which  $p(i)=\sin(i)$ for  $0\leq{}i\leq\pi/2$.
We run the MCMC sampler with 120 walkers for $50\times 10^3$ steps after a burn in phase of $30\times 10^3$ steps. We assessed convergence   through visual inspection of the chains trace plots
and also  by estimating the autocorrelation time \citep{Foreman}, resulting in $\sim$150 steps on average for all  parameters. 
From 
the $6\times10^6$ samples in the MCMC chain, we select as best fit model  the maximum likelihood model, i.e. that with lowest normalised $\chi^2\simeq1$, as given by the following parameters:
$I_0=10.72$\,Jy\,sr$^{-1}$, $R_c=0.46''$, $\gamma_1=-0.39$, $\gamma_2=1.50$, $R_{g,1}=0.12''$, $w_{1g,1}=0.05''$, $w_{2g,1}=0.01''$, $k_{1g,1}=222.04$\,arcsec$^{-1}$, $k_{2g,1}=52.82$\,arcsec$^{-1}$, $\delta_{g,1}=0.98$, $R_{g,2}=0.25''$, $w_{1g,2}=0.29''$, $w_{2g,2}=0.11''$, $k_{1g,2}=95.08$\,arcsec$^{-1}$, $k_{2g,2}=59.77$\,arcsec$^{-1}$, $\delta_{g,2}=0.70$, $R_{g,3}=0.88''$, $w_{1g,3}=0.64''$, $w_{2g,3}=0.06''$, $k_{1g,3}=11.68$\,arcsec$^{-1}$, $k_{2g,3}=21.87$\,arcsec$^{-1}$, $\delta_{g,3}=0.84$, $R_{out}=2.77''$, $i=49.24^\circ$, $PA=11.28^\circ$, $\Delta\mathrm{RA}=0.33''$, $\Delta\mathrm{Dec}=-0.09''$.
This maximum likelihood model falls in the central 68\% interval of the posterior distribution of all  parameters and indeed its brightness profile is representative of the density of models generated by the posterior (see panel (d) of  Figure~\ref{fig:main}). 
 In panel (b) of Figure~\ref{fig:main} we  compare  the observed visibilities and those of the best fit model as a function of deprojected baseline.
  Panel (d) shows a family of $5\times 10^3$ models drawn from the inferred posterior (red lines) and the best fit model (black dot dashed line): the  brightness profile is   tightly constrained between 20 and 100\,au (i.e., the spatial scales probed by most of the interferometric baselines in the dataset) and more uncertain for $R<20$\,au.  
Thus we cannot firmly constrain the detailed shape of the innermost gap, whose width is comparable to the beam ($\sim 7$\,au). We explored in greater detail this degeneracy with a dedicated model suite  and found    an upper limit on the ratio of the flux inside to outside  the gap of  $\sim 0.28$: the gray shaded area in panel (d) highlights the range of brightness values that is compatible with the data. 
 
 The synthesized image of the residuals obtained for the best fit model is shown in panel (c) of Figure~\ref{fig:main}: there are virtually no residuals ($<3\sigma$) in most of the disc at radii $R>25\,$au, confirming the  axisymmetry of the brightness profile. The residuals are most significant  (up to a 12$\sigma$ level)  at the disc center and in the North-West  of the innermost ring. The central residuals reflect the fact that the functional form we have adopted is insufficiently flexible to correctly capture the emissivity profile in the innermost disc.  
 The latter non-axisymmetric residuals might be caused by a combination of optical depth effects owing to the viewing angle of the observations ($i\sim 49^\circ$) and a genuine difference in the local dust temperature.
 
  We note that any perturbation of the disc caused by the hot-Jupiter at 0.1 au would occur on a scale of a few times its orbital radius and would thus be indistinguishable within the synthesized beam of $7\times4$\,au.
 
\section{Modeling the emissivity profile: evidence of multiple planets?}
Structure in protoplanetary discs can derive from many causes.
Non-planetary mechanisms proposed to date are however  not well matched
 to  CI Tau, e.g. photoevaporation produces holes rather than gaps \citep{Ercolano}, while simulations of the vertical shear instability \citep{Flock2}
and of non-ideal MHD effects \citep{Flock1}  do  not
produce the well spaced multiple rings seen in CI Tau. While gaps may also arise from 
opacity effects associated with ice sublimation fronts \citep{Zhang}, the outermost two rings are  well outside the sublimation fronts of even the least  volatile species, N$_2$ and CO \citep{Kwon}.  Moreover, in the innermost gap, our modelling  implies a depletion of the optical depth  by a factor $\sim50$ compared with adjacent regions, considerably more than can be attributed to opacity variations.  
We therefore focus on the planetary hypothesis.
 While gap width can be used to infer the required planet mass
\citep{Rosotti2016}, this conversion depends on
the turbulence in the disc, as parameterised by the Shakura-Sunyaev $\alpha$ parameter \citep{Shakura},
which controls the transport properties of both dust and gas.
The level of disc turbulence is difficult to
constrain observationally: some estimates based on turbulent line broadening have 
suggested very low  $\alpha$ values which then struggle to reproduce 
observed accretion rates \citep{Flaherty} although the universality of this result has been questioned (e.g. \citealt{Teague}).
We explore hydrodynamical models 
in which   $\alpha$ and the gas surface density  are  constrained by  the observed high
accretion rate onto the star ($\dot{M}=3\times10^{-8}M_\odot$ yr$^{-1}$ \citealt{McClure13}), assuming this  
accretion to be driven  by some form of turbulent viscosity; the highly axisymmetric image moreover implies 
that the disc is not self-gravitating.
Together these constraints  favour  a rather high $\alpha$ value ($>10^{-2}$).
We also assume that the maximum grain size,
$a_{max}$ is locally determined by
the minimum of  two limits imposed by radial drift and fragmentation, assuming a fragmentation velocity of $10$ m s $^{-1}$ \citep{Birnstiel}.
We compute
the corresponding dust opacity
\citep{tazzari:2016}
assuming  a population of compact silicate grains with  size distribution  $\mathrm{d}n/\mathrm{d}a\propto{}a^{-3}$ for $a<a_\mathrm{max}$.
and in our emissivity modeling  adopt the temperature profile
derived by \citet{Kwon}. We assume an initial dust to gas ratio of 0.01. 
 
\begin{figure}
\centering
\includegraphics[width=\hsize]{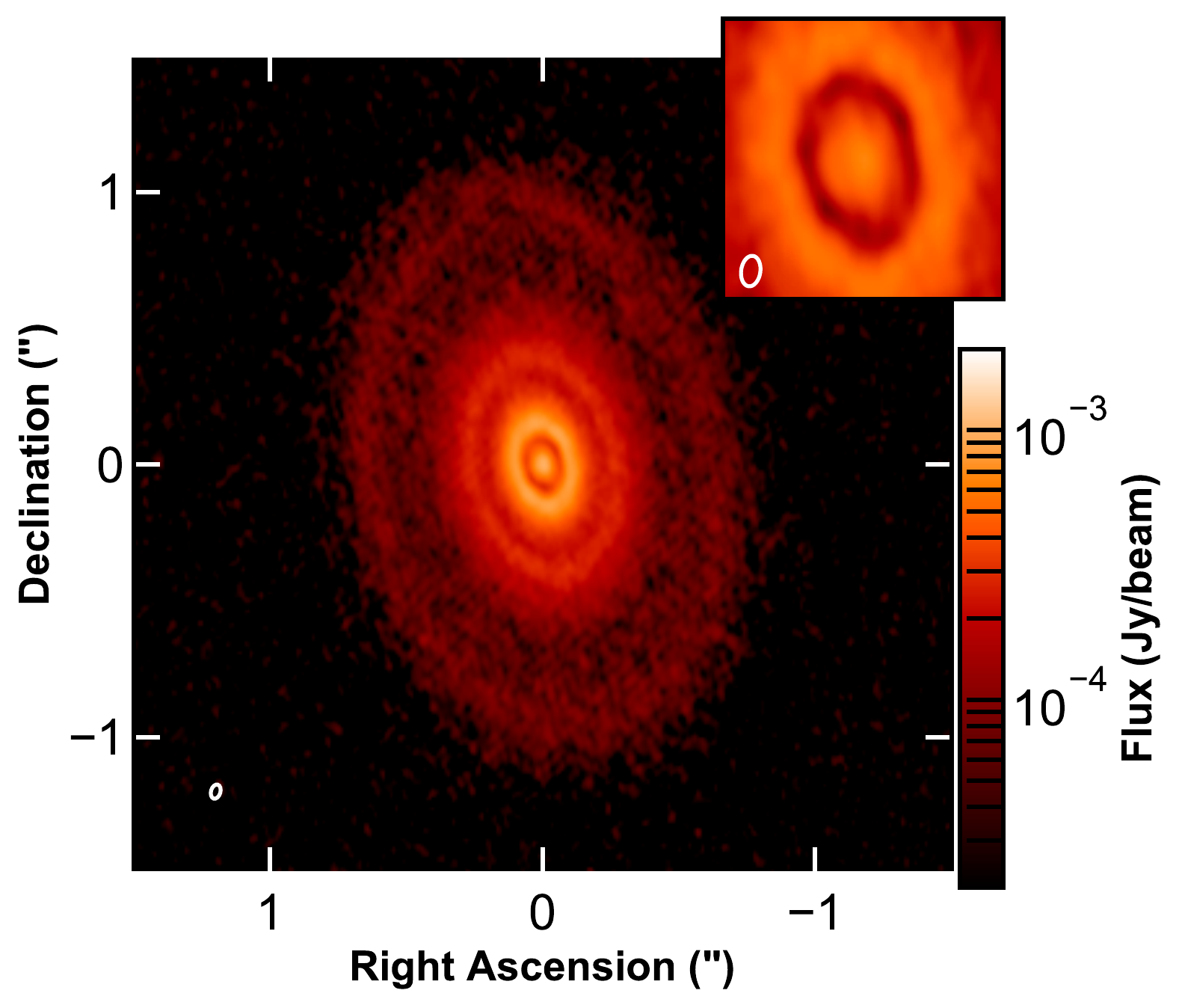}
\caption{CLEANed image of the continuum emission obtained from our gas and dust hydrodynamical simulation containing three planets. This synthetic image was produced with the same noise level as in the observations and using the same imaging parameters used in Fig.~\ref{fig:main}.}
\label{fig:simulated.observations}
\end{figure}

Below we describe simulations of our fiducial model with parameters
$\alpha=0.014$ and total disc mass within
200 au of $20 \Mjupiter$. These parameters imply  strong accretion in the disc ($\sim 10^{-8} M_\odot$ yr$^{-1}$) and yield a profile of $a_{max}$    that is compatible with measurements of the disc-averaged spectral index in CI Tau. {\footnote{A spectral index of $2.9 \pm{} 0.35$ is derived by comparing our $1.3$mm flux  with the $2.7$mm measurements  of 
\citet{Guilloteau2011}.}}
We use these parameters in hydrodynamical  simulations where we insert three planets in the disc,   using the 2D version of the  FARGO3D  code \citep{fargo3d}  with our implementation of drag coupled dust   
\citep{Rosotti2016}.  
We employ 350 logarithmically spaced cells in the radial direction  (from 5.6-378 au)  
and 512 cells  azimuthally, producing approximately square cells at each location. We adopt an  initial gas surface density profile  $\Sigma_\mathrm{gas} \propto 1/r$ normalised at  8\,g\,cm$^{-2}$ at 25\,au, steepening to a $1/r^2$ profile beyond 60\,au.
The local value of $a_{max}$ is computed as above. 
We then compute a synthetic emissivity profile  (using the temperature profile and calculation of opacity as a function of  $a_\mathrm{max}$  detailed above) 
for direct comparison with our GALARIO derived profile.   
The purple dashed line in panel (d) of Figure~\ref{fig:main} presents the brightness profile of our fiducial model where  planets of mass 0.75, 0.15 and 0.4\,$M_\mathrm{Jupiter}$ are located at orbital radii of 14, 43, and 108 au.
We also produce a synthesized image (Fig.~\ref{fig:simulated.observations}), generating model visibilities via the \texttt{ft} task in CASA with exactly the same uv-plane coverage and observational setup as the actual observations and then CLEANing the image using the same imaging parameters as the observed image. 
Note that for a given  disc model, the planet mass within the innermost gap is only determined to within around a factor of two since  this gap is poorly resolved, while the values are constrained to within around $30\%$ in the outer two gaps.
\section{Discussion}
\subsection{Observational tests of the fiducial model}
Our fiducial model is motivated by reproducing accretion rate and spectral index data for CI Tau, which results in moderately high turbulence levels ($\alpha \sim 0.01$).
In HL Tau \citet{Pinte} have argued for low turbulence levels on account of  the narrowness of the ring features; in CI Tau, however,  the somewhat wider gaps and colder disc means that the turbulence levels cannot be constrained in this way. The ratio of total fluxes at $2.7$ to $1.3$\,mm in fact {\it requires} that $a_{max}$ in the outer, optically thin, regions of CI Tau is relatively low ($<1$\,mm), in agreement with our model. CI Tau may be relatively unusual in lacking larger grains
(its mm  spectral index  lies at  the $\sim 85$th percentile among protoplanetary discs; \citealt{Testi}). 
    An alternative scenario, if we put aside the evidence from the mm spectral index for small grains in the outer disc,  is that disc
accretion is driven by  a magnetised wind (e.g., \citealt{Bai}).
rather than  turbulent viscosity. Low turbulence levels allow grains to grow and partially decouple from the flow so that lower planetary masses are required to match the observed gap parameters:   from the
hydrodynamical simulations of \citet{Rosotti2016}  (where $\alpha = 10^{-3}$) we estimate planet masses of 20-30 earth masses for the outer two 
although a mass of up to around a  Jupiter mass can be accommodated in the case of the
innermost planet. Spatially resolved spectral index determinations
\citep{tazzari:2016}
as well as searches for possible kinematic distortions expected from a
    gas giant planet \citep{Pinte18,Teague18}, could potentially discriminate between these possibilities.

 \subsection{Evolutionary scenarios for the fiducial model: formation and migration}
 
 The inferred planet masses in the three gaps suggest that none
of these planets formed through gravitational instability. Planets
formed in this way should exceed the Jeans limit in the outer disc
(about a  Jupiter mass) and should rapidly grow to much larger
masses by accretion \citep{Kratter}. The hot Jupiter on the other
hand could have been formed by a variety  of mechanisms; from the
modeled masses in disc and planets and from the accretion on to the
star the inferred timescale for its inward migration is $\sim 0.4$ Myr \citep{Duermann} so that there would have been plenty of time for it
to have migrated from a range of outward lying locations.
The roughly Jovian mass planet inferred at $14$ au is also  easy to
account for in terms of existing planet formation models (i.e.  core
accretion models involving either planetesimal or pebble accretion  
\citep{ida,bitsch}). However neither of these models readily account for
the two lower mass planets  
at $43$ and
$108$ au. The timescales for forming and accumulating solid material
are long in the outer disc (though see \citealt{Rafikov:2011aa}  for arguments
in favour of planetesimal accretion at large orbital radii). Even if this
were circumvented, these planets would have had to have grown through the
mass range (10-20 earth masses) where
rapid inward migration is expected \citep{Paardekooper}
 and so their existence at large radii
is a puzzle. \citet{ida} were able to generate a modest population of gas giants at large separations through outward scattering of planetary cores and subsequent accretion but their population synthesis models only sparsely populate the parameter space corresponding to the two outer planets in CI Tau.

 It is unclear whether the current planetary architecture would survive on Gyr timescales. The planets' period ratios do not suggest a resonant configuration. Nevertheless, the relatively high disc mass means that they may still   end up at small radii, possibly being swallowed by the star or ejected from the system by scattering off the hot Jupiter \citep{Lega}.   
 While current imaging surveys of mature systems do not have the sensitivity to detect planets of the masses we infer in CI Tau \citep{Vigan}, future surveys will be able to determine if CI Tau-like systems are long lived.
 
\subsection{Comparison with other gapped discs}
High resolution ALMA studies are steadily increasing the census of discs with annular substructure. Although unique in being the only such system with a hot
Jupiter, CI Tau's ring structure is not unusual. Its well spaced broad annuli place it in a similar category to HL Tau, HD 163296 and HD 169142 \citep{ALMAP,Isella,Fedele17}; none of the above  systems exhibit the closely spaced shallow features seen in TW Hydra \citep{Andrews16} or the narrow deep features seen in AS 209 \citep{Fedele}. However only TW Hydra and HL Tau have been observed at a comparably high resolution to   this study.
Previous modeling of the gap structures in HL Tau  (e.g. \citealt{Dipierro}) have yielded   similar  planet masses as a function of disc parameters to what we   report here, although in HL Tau the choice of disc model has not been  constrained by   other system observables. 
\section{Conclusions}
 High resolution ALMA data of the disc in the young star CI Tau has revealed three prominent  annular emission gaps  which we have interpreted as  an ensemble of massive planets   spanning a factor thousand in orbital radius. 
 The wealth of supplementary data available on CI Tau  has allowed us to construct models that are consistent with {\it all} the data on this system available to date.  The inferred planetary architecture  suggests that the observed association between hot Jupiter and other companions may be in place at very early times.
  We note that the outer two planets (sub Jovian planets at radii of $43$ and $108$ au) present a challenge to current planet formation models.
\acknowledgments{
This work is supported by the DISCSIM project, grant agreement 341137 funded by the European Research Council under ERC-2013-ADG. MK acknowledges funding from the European Union'€™s Horizon 2020 programme, Marie Sklodowska-Curie grant
agreement no. 753799 and  FM from a Leverhulme Trust Early Career Fellowship,the Isaac Newton Trust and a Royal Society Dorothy Hodgkin Fellowship.  This paper  uses  ALMA data ADS/JAO.ALMA\#2016.1.01370.S. ALMA is a partnership of ESO (representing its member states), NSF (USA) and NINS (Japan), together with NRC (Canada) and NSC and ASIAA (Taiwan) and KASI (Republic of Korea), in cooperation with the Republic of Chile. The Joint ALMA Observatory is operated by ESO, AUI/NRAO and NAOJ.}
\software{GALARIO v1.2 \citep{Tazzari}, emcee v2.2.0 \citep{Foreman}, CASA v5.1.1 \citep{McMullin:2007aa}.}
\bibliographystyle{aasjournal}
\bibliography{ci_tau_refs}

\begin{thebibliography}{}
\expandafter\ifx\csname natexlab\endcsname\relax\def\natexlab#1{#1}\fi
\providecommand{\url}[1]{\href{#1}{#1}}

\bibitem[{{ALMA Partnership} {et~al.}(2015){ALMA Partnership}, {Brogan},
  {P{\'e}rez}, {Hunter}, {Dent}, {Hales}, {Hills}, {Corder}, {Fomalont},
  {Vlahakis}, {Asaki}, {Barkats}, {Hirota}, {Hodge}, {Impellizzeri}, {Kneissl},
  {Liuzzo}, {Lucas}, {Marcelino}, {Matsushita}, {Nakanishi}, {Phillips},
  {Richards}, {Toledo}, {Aladro}, {Broguiere}, {Cortes}, {Cortes}, {Espada},
  {Galarza}, {Garcia-Appadoo}, {Guzman-Ramirez}, {Humphreys}, {Jung}, {Kameno},
  {Laing}, {Leon}, {Marconi}, {Mignano}, {Nikolic}, {Nyman}, {Radiszcz},
  {Remijan}, {Rod{\'o}n}, {Sawada}, {Takahashi}, {Tilanus}, {Vila Vilaro},
  {Watson}, {Wiklind}, {Akiyama}, {Chapillon}, {de Gregorio-Monsalvo}, {Di
  Francesco}, {Gueth}, {Kawamura}, {Lee}, {Nguyen Luong}, {Mangum}, {Pietu},
  {Sanhueza}, {Saigo}, {Takakuwa}, {Ubach}, {van Kempen}, {Wootten},
  {Castro-Carrizo}, {Francke}, {Gallardo}, {Garcia}, {Gonzalez}, {Hill},
  {Kaminski}, {Kurono}, {Liu}, {Lopez}, {Morales}, {Plarre}, {Schieven},
  {Testi}, {Videla}, {Villard}, {Andreani}, {Hibbard}, \& {Tatematsu}}]{ALMAP}
{ALMA Partnership}, {Brogan}, C.~L., {P{\'e}rez}, L.~M., {et~al.} 2015, \apjl,
  808, L3

\bibitem[{{Andrews} \& {Williams}(2007)}]{Andrews07}
{Andrews}, S.~M., \& {Williams}, J.~P. 2007, \apj, 659, 705

\bibitem[{{Andrews} {et~al.}(2016){Andrews}, {Wilner}, {Zhu}, {Birnstiel},
  {Carpenter}, {P{\'e}rez}, {Bai}, {{\"O}berg}, {Hughes}, {Isella}, \&
  {Ricci}}]{Andrews16}
{Andrews}, S.~M., {Wilner}, D.~J., {Zhu}, Z., {et~al.} 2016, \apjl, 820, L40

\bibitem[{{Bai}(2016)}]{Bai}
{Bai}, X.-N. 2016, \apj, 821, 80

\bibitem[{{Becker} {et~al.}(2017){Becker}, {Vanderburg}, {Adams}, {Khain}, \&
  {Bryan}}]{Becker}
{Becker}, J.~C., {Vanderburg}, A., {Adams}, F.~C., {Khain}, T., \& {Bryan}, M.
  2017, \aj, 154, 230

\bibitem[{{Ben{\'{\i}}tez-Llambay} \& {Masset}(2016)}]{fargo3d}
{Ben{\'{\i}}tez-Llambay}, P., \& {Masset}, F.~S. 2016, \apjs, 223, 11

\bibitem[{{Biddle} {et~al.}(2018){Biddle}, {Johns-Krull}, {Llama}, {Prato}, \&
  {Skiff}}]{Biddle}
{Biddle}, L.~I., {Johns-Krull}, C.~M., {Llama}, J., {Prato}, L., \& {Skiff},
  B.~A. 2018, \apjl, 853, L34

\bibitem[{{Birnstiel} {et~al.}(2012){Birnstiel}, {Klahr}, \&
  {Ercolano}}]{Birnstiel}
{Birnstiel}, T., {Klahr}, H., \& {Ercolano}, B. 2012, \aap, 539, A148

\bibitem[{{Bitsch} {et~al.}(2015){Bitsch}, {Lambrechts}, \&
  {Johansen}}]{bitsch}
{Bitsch}, B., {Lambrechts}, M., \& {Johansen}, A. 2015, \aap, 582, A112

\bibitem[{{Dipierro} {et~al.}(2015){Dipierro}, {Price}, {Laibe}, {Hirsh},
  {Cerioli}, \& {Lodato}}]{Dipierro}
{Dipierro}, G., {Price}, D., {Laibe}, G., {et~al.} 2015, \mnras, 453, L73

\bibitem[{{D{\"u}rmann} \& {Kley}(2015)}]{Duermann}
{D{\"u}rmann}, C., \& {Kley}, W. 2015, \aap, 574, A52

\bibitem[{{Ercolano} {et~al.}(2017){Ercolano}, {Rosotti}, {Picogna}, \&
  {Testi}}]{Ercolano}
{Ercolano}, B., {Rosotti}, G.~P., {Picogna}, G., \& {Testi}, L. 2017, \mnras,
  464, L95

\bibitem[{{Fedele} {et~al.}(2017){Fedele}, {Carney}, {Hogerheijde}, {Walsh},
  {Miotello}, {Klaassen}, {Bruderer}, {Henning}, \& {van Dishoeck}}]{Fedele17}
{Fedele}, D., {Carney}, M., {Hogerheijde}, M.~R., {et~al.} 2017, \aap, 600, A72

\bibitem[{{Fedele} {et~al.}(2018){Fedele}, {Tazzari}, {Booth}, {Testi},
  {Clarke}, {Pascucci}, {Kospal}, {Semenov}, {Bruderer}, {Henning}, \&
  {Teague}}]{Fedele}
{Fedele}, D., {Tazzari}, M., {Booth}, R., {et~al.} 2018, \aap, 610, A24

\bibitem[{{Flaherty} {et~al.}(2015){Flaherty}, {Hughes}, {Rosenfeld},
  {Andrews}, {Chiang}, {Simon}, {Kerzner}, \& {Wilner}}]{Flaherty}
{Flaherty}, K.~M., {Hughes}, A.~M., {Rosenfeld}, K.~A., {et~al.} 2015, \apj,
  813, 99

\bibitem[{{Flock} {et~al.}(2017){Flock}, {Nelson}, {Turner}, {Bertrang},
  {Carrasco-Gonz{\'a}lez}, {Henning}, {Lyra}, \& {Teague}}]{Flock2}
{Flock}, M., {Nelson}, R.~P., {Turner}, N.~J., {et~al.} 2017, \apj, 850, 131

\bibitem[{{Flock} {et~al.}(2015){Flock}, {Ruge}, {Dzyurkevich}, {Henning},
  {Klahr}, \& {Wolf}}]{Flock1}
{Flock}, M., {Ruge}, J.~P., {Dzyurkevich}, N., {et~al.} 2015, \aap, 574, A68

\bibitem[{{Foreman-Mackey} {et~al.}(2013){Foreman-Mackey}, {Hogg}, {Lang}, \&
  {Goodman}}]{Foreman}
{Foreman-Mackey}, D., {Hogg}, D.~W., {Lang}, D., \& {Goodman}, J. 2013, \pasp,
  125, 306

\bibitem[{{Guilloteau} {et~al.}(2011){Guilloteau}, {Dutrey}, {Pi{\'e}tu}, \&
  {Boehler}}]{Guilloteau2011}
{Guilloteau}, S., {Dutrey}, A., {Pi{\'e}tu}, V., \& {Boehler}, Y. 2011, \aap,
  529, A105

\bibitem[{{Guilloteau} {et~al.}(2014){Guilloteau}, {Simon}, {Pi{\'e}tu}, {Di
  Folco}, {Dutrey}, {Prato}, \& {Chapillon}}]{Guilloteau14}
{Guilloteau}, S., {Simon}, M., {Pi{\'e}tu}, V., {et~al.} 2014, \aap, 567, A117

\bibitem[{{Haisch} {et~al.}(2001){Haisch}, {Lada}, \& {Lada}}]{Haisch}
{Haisch}, Jr., K.~E., {Lada}, E.~A., \& {Lada}, C.~J. 2001, \apjl, 553, L153

\bibitem[{{Ida} {et~al.}(2013){Ida}, {Lin}, \& {Nagasawa}}]{ida}
{Ida}, S., {Lin}, D.~N.~C., \& {Nagasawa}, M. 2013, \apj, 775, 42

\bibitem[{Isella {et~al.}(2016)Isella, Guidi, Testi, Liu, Li, Li, Weaver,
  Boehler, Carperter, De~Gregorio-Monsalvo, Manara, Natta, P\'erez, Ricci,
  Sargent, Tazzari, \& Turner}]{Isella}
Isella, A., Guidi, G., Testi, L., {et~al.} 2016, Phys. Rev. Lett., 117, 251101

\bibitem[{{Johns-Krull} {et~al.}(2016){Johns-Krull}, {McLane}, {Prato},
  {Crockett}, {Jaffe}, {Hartigan}, {Beichman}, {Mahmud}, {Chen}, {Skiff},
  {Cauley}, {Jones}, \& {Mace}}]{Johns-Krull}
{Johns-Krull}, C.~M., {McLane}, J.~N., {Prato}, L., {et~al.} 2016, \apj, 826,
  206

\bibitem[{{Kley} \& {Nelson}(2012)}]{Kley}
{Kley}, W., \& {Nelson}, R.~P. 2012, \araa, 50, 211

\bibitem[{{Knutson} {et~al.}(2014){Knutson}, {Fulton}, {Montet}, {Kao}, {Ngo},
  {Howard}, {Crepp}, {Hinkley}, {Bakos}, {Batygin}, {Johnson}, {Morton}, \&
  {Muirhead}}]{Knutson2014}
{Knutson}, H.~A., {Fulton}, B.~J., {Montet}, B.~T., {et~al.} 2014, \apj, 785,
  126

\bibitem[{{Konishi} {et~al.}(2018){Konishi}, {Hashimoto}, \&
  {Hori}}]{Konishi2018}
{Konishi}, M., {Hashimoto}, J., \& {Hori}, Y. 2018, \apjl, 859, L28

\bibitem[{{Kratter} \& {Lodato}(2016)}]{Kratter}
{Kratter}, K., \& {Lodato}, G. 2016, \araa, 54, 271

\bibitem[{{Kwon} {et~al.}(2015){Kwon}, {Looney}, {Mundy}, \& {Welch}}]{Kwon}
{Kwon}, W., {Looney}, L.~W., {Mundy}, L.~G., \& {Welch}, W.~J. 2015, \apj, 808,
  102

\bibitem[{{Lega} {et~al.}(2013){Lega}, {Morbidelli}, \& {Nesvorn{\'y}}}]{Lega}
{Lega}, E., {Morbidelli}, A., \& {Nesvorn{\'y}}, D. 2013, \mnras, 431, 3494

\bibitem[{{Mayor} \& {Queloz}(1995)}]{Mayor}
{Mayor}, M., \& {Queloz}, D. 1995, \nat, 378, 355

\bibitem[{{McClure} {et~al.}(2013){McClure}, {Calvet}, {Espaillat}, {Hartmann},
  {Hern{\'a}ndez}, {Ingleby}, {Luhman}, {D'Alessio}, \& {Sargent}}]{McClure13}
{McClure}, M.~K., {Calvet}, N., {Espaillat}, C., {et~al.} 2013, \apj, 769, 73

\bibitem[{{McMullin} {et~al.}(2007){McMullin}, {Waters}, {Schiebel}, {Young},
  \& {Golap}}]{McMullin:2007aa}
{McMullin}, J.~P., {Waters}, B., {Schiebel}, D., {Young}, W., \& {Golap}, K.
  2007, in Astronomical Society of the Pacific Conference Series, Vol. 376,
  Astronomical Data Analysis Software and Systems XVI, ed. R.~A. {Shaw},
  F.~{Hill}, \& D.~J. {Bell}, 127

\bibitem[{{Ngo} {et~al.}(2015){Ngo}, {Knutson}, {Hinkley}, {Crepp}, {Bechter},
  {Batygin}, {Howard}, {Johnson}, {Morton}, \& {Muirhead}}]{Ngo2015}
{Ngo}, H., {Knutson}, H.~A., {Hinkley}, S., {et~al.} 2015, \apj, 800, 138

\bibitem[{{Paardekooper} {et~al.}(2011){Paardekooper}, {Baruteau}, \&
  {Kley}}]{Paardekooper}
{Paardekooper}, S.-J., {Baruteau}, C., \& {Kley}, W. 2011, \mnras, 410, 293

\bibitem[{{Pinte} {et~al.}(2016){Pinte}, {Dent}, {M{\'e}nard}, {Hales}, {Hill},
  {Cortes}, \& {de Gregorio-Monsalvo}}]{Pinte}
{Pinte}, C., {Dent}, W.~R.~F., {M{\'e}nard}, F., {et~al.} 2016, \apj, 816, 25

\bibitem[{{Pinte} {et~al.}(2018){Pinte}, {Price}, {M{\'e}nard}, {Duch{\^e}ne},
  {Dent}, {Hill}, {de Gregorio-Monsalvo}, {Hales}, \& {Mentiplay}}]{Pinte18}
{Pinte}, C., {Price}, D.~J., {M{\'e}nard}, F., {et~al.} 2018, \apjl, 860, L13

\bibitem[{{Rafikov}(2011)}]{Rafikov:2011aa}
{Rafikov}, R.~R. 2011, \apj, 727, 86

\bibitem[{{Rasio} \& {Ford}(1996)}]{Rasio}
{Rasio}, F.~A., \& {Ford}, E.~B. 1996, Science, 274, 954

\bibitem[{{Rosotti} {et~al.}(2016){Rosotti}, {Juhasz}, {Booth}, \&
  {Clarke}}]{Rosotti2016}
{Rosotti}, G.~P., {Juhasz}, A., {Booth}, R.~A., \& {Clarke}, C.~J. 2016,
  \mnras, 459, 2790

\bibitem[{{Shakura} \& {Sunyaev}(1973)}]{Shakura}
{Shakura}, N.~I., \& {Sunyaev}, R.~A. 1973, \aap, 24, 337

\bibitem[{{Simon} {et~al.}(2017){Simon}, {Guilloteau}, {Di Folco}, {Dutrey},
  {Grosso}, {Pi{\'e}tu}, {Chapillon}, {Prato}, {Schaefer}, {Rice}, \&
  {Boehler}}]{Simon17}
{Simon}, M., {Guilloteau}, S., {Di Folco}, E., {et~al.} 2017, \apj, 844, 158

\bibitem[{{Tazzari} {et~al.}(2018){Tazzari}, {Beaujean}, \& {Testi}}]{Tazzari}
{Tazzari}, M., {Beaujean}, F., \& {Testi}, L. 2018, \mnras, 476, 4527

\bibitem[{{Tazzari} {et~al.}(2016){Tazzari}, {Testi}, {Ercolano}, {Natta},
  {Isella}, {Chandler}, {P{\'e}rez}, {Andrews}, {Wilner}, {Ricci}, {Henning},
  {Linz}, {Kwon}, {Corder}, {Dullemond}, {Carpenter}, {Sargent}, {Mundy},
  {Storm}, {Calvet}, {Greaves}, {Lazio}, \& {Deller}}]{tazzari:2016}
{Tazzari}, M., {Testi}, L., {Ercolano}, B., {et~al.} 2016, \aap, 588, A53

\bibitem[{{Teague} {et~al.}(2018){Teague}, {Bae}, {Bergin}, {Birnstiel}, \&
  {Foreman-Mackey}}]{Teague18}
{Teague}, R., {Bae}, J., {Bergin}, E.~A., {Birnstiel}, T., \& {Foreman-Mackey},
  D. 2018, \apjl, 860, L12

\bibitem[{{Teague} {et~al.}(2016){Teague}, {Guilloteau}, {Semenov}, {Henning},
  {Dutrey}, {Pi{\'e}tu}, {Birnstiel}, {Chapillon}, {Hollenbach}, \&
  {Gorti}}]{Teague}
{Teague}, R., {Guilloteau}, S., {Semenov}, D., {et~al.} 2016, \aap, 592, A49

\bibitem[{{Testi} {et~al.}(2014){Testi}, {Birnstiel}, {Ricci}, {Andrews},
  {Blum}, {Carpenter}, {Dominik}, {Isella}, {Natta}, {Williams}, \&
  {Wilner}}]{Testi}
{Testi}, L., {Birnstiel}, T., {Ricci}, L., {et~al.} 2014, Protostars and
  Planets VI, 339

\bibitem[{{Vigan} {et~al.}(2017){Vigan}, {Bonavita}, {Biller}, {Forgan},
  {Rice}, {Chauvin}, {Desidera}, {Meunier}, {Delorme}, {Schlieder}, {Bonnefoy},
  {Carson}, {Covino}, {Hagelberg}, {Henning}, {Janson}, {Lagrange}, {Quanz},
  {Zurlo}, {Beuzit}, {Boccaletti}, {Buenzli}, {Feldt}, {Girard}, {Gratton},
  {Kasper}, {Le Coroller}, {Mesa}, {Messina}, {Meyer}, {Montagnier},
  {Mordasini}, {Mouillet}, {Moutou}, {Reggiani}, {Segransan}, \&
  {Thalmann}}]{Vigan}
{Vigan}, A., {Bonavita}, M., {Biller}, B., {et~al.} 2017, \aap, 603, A3

\bibitem[{{Wright} {et~al.}(2012){Wright}, {Marcy}, {Howard}, {Johnson},
  {Morton}, \& {Fischer}}]{Wright}
{Wright}, J.~T., {Marcy}, G.~W., {Howard}, A.~W., {et~al.} 2012, \apj, 753, 160

\bibitem[{{Zhang} {et~al.}(2016){Zhang}, {Bergin}, {Blake}, {Cleeves},
  {Hogerheijde}, {Salinas}, \& {Schwarz}}]{Zhang}
{Zhang}, K., {Bergin}, E.~A., {Blake}, G.~A., {et~al.} 2016, \apjl, 818, L16

\end{thebibliography}
\end{document}